\documentclass[a4paper,11pt]{article}
\usepackage{pos}
\usepackage{abbrv}

\title{Neutrino Imaging of the Galactic Centre 
  and Millisecond Pulsar Population}

\author*[a]{Paul C.~W.~Lai}
\author[b]{Matteo Agostini}
\author[c]{Foteini Oikonomou}
\author[b]{Beatrice Crudele}
\author[d]{Ellis R.~Owen}
\author[a]{Kinwah Wu}

\affiliation[a]{Mullard Space Science Laboratory, University College London, \\Holmbury St.~Mary, Surrey RH5 6NT, United Kingdom}

\affiliation[b]{Department of Physics and Astronomy, University College London, \\
 Gower Street, London, WC1E 6BT, United Kingdom}

\affiliation[c]{Institutt for Fysikk, Norwegian University of Science and Technology, \\ Trondheim, Norway}

\affiliation[d]{Theoretical Astrophysics, Department of Earth and Space Science, Graduate School of Science, \\ Osaka University, Toyonaka,
Osaka 560-0043, Japan}

\emailAdd{chong.lai.22@ucl.ac.uk}

\abstract{
In this work, we consider the possible presence of a large population of millisecond pulsars in the Galactic Centre. Their direct detection would be challenging due to severe pulse broadening caused by scattering of radiation. 
We propose a new method to constrain 
    their population with neutrino imaging
    of the Galactic Centre.
Millisecond pulsars are proposed cosmic-ray accelerators. The high-energy protons they produce will collide with the baryonic matter in the central molecular zone to create charged and neutral pions that decay into neutrinos and $\gamma$-rays, respectively.  
The specific neutrino and $\gamma$-ray fluxes   
  must be below their corresponding observed values, allowing us to put a conservative upper limit on the millisecond pulsar population 
    of $N_{\rm MSP} < 10{,}000$ within a galacto-centric radius of 20\,pc. 
This upper limit is sensitive to
   the proton acceleration efficiency 
   of the pulsars, but is less dependent on
    the particle injection spectral index and
    the choice of mass tracers.
The population will be better constrained
    when high resolution
    neutrino observations
    of the Galactic Centre become
    available. 
The presence of these millisecond pulsars 
  can account for the $\gamma$-ray excess 
  in the Galactic Centre.
}

\ConferenceLogo{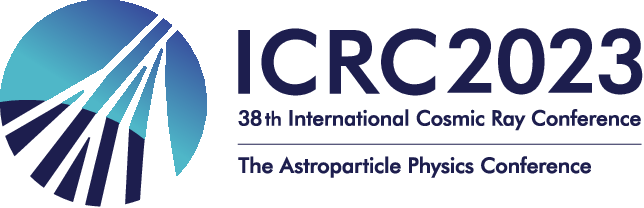}

\FullConference{%
38th International Cosmic Ray Conference (ICRC2023)\\
  26 July - 3 August, 2023\\
  Nagoya, Japan}

\begin{document}
\maketitle

\section{Overview} 

It has been suggested that the Galactic Centre (GC) 
    habours a large number of pulsars 
  \citep[see][]{Pfahl2004ApJ,Wharton12ApJ}. 
Verifying their presence or absence  
  will help to resolve issues 
  such as the origin(s) of the multiple components of  $\gamma$-rays, 
  and the stellar-mass black-hole content in the GC region.    
The fast rotation and strong magnetic field 
  of a pulsar 
  create a large electric potential which
  would accelerate charged particles to
  high energies.
There is clear evidence  
    showing pulsars are able to produce 
    high-energy electrons and positrons
        \citep[e.g. from studies of the Crab Pulsar,][]{Buhler+14}.
Accelerated high energy protons emit radiation much less strongly than their leptonic counterparts. 
    Evidence for their presence is therefore  more
    challenging to establish, and an unambiguous observational
    signature of hadronic particles from pulsars is yet to be found. 
However, 
the acceleration of hadronic particles in pulsar environments has not been 
ruled-out, and 
    theoretical studies have 
    considered the possibility that 
    pulsars can accelerate both protons
    and leptons 
    \citep[e.g.][]{Blasi00ApJL, Arons03ApJ, Guepin20A&A}. 
    This study therefore also introduces an indirect
    means to reveal the 
    presence of cosmic-ray (CR) protons accelerated by pulsar populations. 


The TeV $\gamma$-ray emission observed from the GC region  
  implies the presence of localised PeV accelerators, 
  if the $\gamma$-rays are by-products of CR interactions 
  \citep[see][]{HESS16Nat, HESS18A&A}.  
The massive nuclear black hole (Sgr A*), 
    supernova remnants, 
   star clusters and pulsars are   
   candidate PeV charged-particle accelerators.
Considering a CR diffusion timescale of 
     $t_{\rm diff} = r^2/(4D) \sim 3\times10^5\,$yr
    for $r \sim 200\,$pc and 
    $D \sim 10^{28}\,$cm$^2$\,s$^{-1}$,
    the $1/r$ spatial distribution of CR density in the Galaxy 
    suggests that 
    Sgr A* must have been active
    until at least 
    $3\times10^5$\,yr ago, if it were the primary source of CRs \citep{HESS16Nat, MAGIC20A&A}. 
Sgr A* has experienced several outbursts over 
    the past few hundred years 
    \citep[see e.g.][]{Rogers+22}, however it has spent 
    most of this time in a 
    quiescent state. 
    The energy output of the recent outbursts 
    is not sufficient to account for 
    the abundance of GC CRs.
If associated with Sgr A*, the estimated age range of the \textit{Fermi} bubbles indicates that it  underwent an 
 active episode 
     1-10\,Myr ago
    \citep{Guo12ApJ, Yang22NatAst}. 
This episode may also be 
    insufficient to produce the 
    observed Galactic CR distribution.  
Indeed, H.E.S.S. observations \citep{HESS16Nat}
   suggest that 
   the injection of PeV CRs at the GC 
   is more likely to be continuous 
    rather than episodic. 
Extreme supernova remnants could produce PeV CRs 
  \cite{Marcowith18MNRAS}, 
  but such extreme supernovae are uncommon.  
The two remaining CR source candidates  
  are therefore nuclear star clusters and pulsars, in particular millisecond pulsars (MSPs) 
    \cite{Guepin20A&A, Morlino21MNRAS}. 
In this work, we focus on the latter case. 
  
Direct detection of GC MSPs  
  through traditional timing radio observations 
  is difficult. 
  This is because of 
  severe pulse broadening caused by scattering
  in 
  ionised gas. 
Alternative means must therefore be sought 
  to detect this pulsar population, if present. 
We propose that neutrino imaging of the 
    GC 
    will constrain the 
    MSP population 
    more stringently 
    than the use of 
    $\gamma$-rays.
The recent IceCube detection of the Milky Way neutrino flux 
    has an angular resolution of $\sim 10^\circ$ \citep{IceCube2023Sci}, much larger than 
    the region of interest in this work ($\sim 1^\circ$). 
    The IceCube result therefore cannot be directly compared 
    to the results presented in this study.
However, 
higher-fidelity 
  neutrino observations will be achievable in the near future, 
  and it is timely to discuss the applications of neutrino imaging.
Future observations will be able to constrain
  the GC MSP population, and hence shed light on a number of 
  open questions about the GC region, including 
  its $\gamma$-ray excess and the missing pulsar
  problem \citep[see e.g.][]{Wharton12ApJ, Bartels16PRL}. 

\section{Model Scenario} 

The scenario we consider in this work is summarised as follows. 
A large number of MSPs reside in the
    stellar cluster around Sgr A*.
Protons accelerated by these MSPs are scattered by the interstellar
 magnetic field. 
Co-located with the MSPs in the GC 
  is a spatially extended diffuse baryonic medium,
  where most of the baryons are locked in 
  the clouds of the central molecular zone (CMZ). 
The GC CRs interact with the CMZ baryons
  via p-p collisions. 
  These interactions produce charged and neutral pions
  which rapidly decay into neutrinos and $\gamma$-rays,
  respectively.


\subsection{Millisecond pulsars as particle accelerators}  

For a MSP with a rotational period $P_{\rm s}$ 
   and polar strength $B$ for a dipolar magnetic field, 
   the spin-down luminosity is  
\begin{align} 
    L_{\rm sd} & 
    = 1.4\times10^{35}\,
    {\rm erg}\,{\rm s}^{-1}\,
    \left(
    \frac{B}{10^8\,{\rm G}} 
    \right)^2
    \left(
    \frac{R_{\rm ns}}{10\,{\rm km}}
    \right)^6
    \left(
    \frac{P_{\rm s}}{1\,{\rm ms}}
    \right)^{-4} \ ,
\label{eq:spin_down_lum}
\end{align}   
  where $R_{\rm ns}$ is the MSP (neutron star) radius. 
The maximum energy can be attained by a proton is 
\begin{align}
    E_{\rm p, max} &  = 2\;\! \eta\,
    {\rm PeV}\,
    \left(
    \frac{B}{10^8\,{\rm G}} 
    \right)
    \left(
    \frac{R_{\rm ns}}{10\,{\rm km}}
    \right)^2
    \left(
    \frac{P_{\rm s}}{1\,{\rm ms}}
    \right)^{-1} \ 
\label{eq:p_acceleration} 
\end{align} 
\cite{Guepin20A&A}.
The maximum energy depends on the 
    pair production efficiency.
The higher the efficiency, 
    the lower the maximum energy.
This is parametrised by
    $\eta$, which ranges from 0.25 to 1.
As only a fraction of the rotational power of a MSP  
  is available for proton acceleration,  
  a scaling parameter $f_{\rm p}$ is introduced.  
This gives the CR power of a MSP:     
\begin{align}
    f_{\rm p}\;\!L_{\rm sd} 
    &  = 
    \int^\infty_{m_{\rm p}c^2} 
    {\rm d}E_{\rm p}  \  
    Q_{\rm psr}(E_{\rm p})\;\! E_{\rm p}  
    \propto 
    \int^\infty_{m_{\rm p}c^2} 
    {\rm d}E_{\rm p}  \  
    E^{\Gamma+1} e^{-E_{\rm p}/E_{\rm p, max}} \ , 
\label{eq:injection}
\end{align}    
 where $Q_{\rm psr}(E_{\rm p})$ 
   is the spectral power in units of $({\rm erg~ s})^{-1}$,    
    $\Gamma$ is spectral index 
     of CR protons accelerated by the MSP, 
     and $m_{\rm p}$ is the proton rest mass. 
The high-energy spectral cut-off 
  is set by $E_{\rm p,max}$ 
  (see Equation~\ref{eq:p_acceleration}).

\subsection{Cosmic-ray transport with hadronic 
 interactions}  

We model the distribution of CR protons
  using the transport equation in the diffusion limit, characterised 
  by an energy-dependent diffusion coefficient, $D(E_{\rm p})$: 
\begin{align}
    \frac{\partial n_{\rm CRp}(E_{\rm p}, r, t)}{\partial t} 
  -      \nabla \cdot \left[  
  D(E_{\rm p}) \nabla n_{\rm CRp}(E_{\rm p}, r) \right]
   & = Q_{\rm psr}(E_{\rm p})\;\! n_{\rm MSP}(r) 
    - c \left[  
    n_{\rm H}\;\! \sigma_{\rm pp}(E_{\rm p})  \right] 
    n_{\rm CRp}(E_{\rm p}, r, t)  
    \ , 
\label{eq:transport_eq}
\end{align}
  where   
   $n_{\rm CRp}$ is the CR proton number density, 
   $n_{\rm H}$ is the Hydrogen number density in the diffuse medium, 
    and $n_{\rm MSP}$ is the MSP number density in the GC region.  
The two terms on the right side of the equation 
   are the rates of particle loss 
   (through inelastic hadronic p-p interactions 
   with a cross-section $\sigma_{\rm pp}$)   
   and of production of CR protons by MSPs.   
The solution to the transport equation is 
\begin{align} 
\label{eq:solution}
    n_{\rm CRp}(E_{\rm p}, r) &  \approx
    \frac{Q_{\rm psr}(E_{\rm p})}{4\pi D(E_{\rm p})}
    \int_r^\infty {\rm d} r' \ 
    \frac{N_{\rm MSP}(r')}{r'^2} \ 
\end{align}  
  in the limits $r \ll 2\sqrt{Dt}$ 
  (stationary limit)
  and 
    $r \ll \sqrt{D/(cn_{\rm H}\sigma_{\rm pp})}$
    (negligible absorption),  
 with the number of pulsars within a radius $r'$ given by 
$N_{\rm MSP}(r') = \int_0^{r'}{\rm d}r\;\!
    n_{\rm MSP}(r)\;\! 4\pi r^2$. 
The all-favour neutrino  
  and $\gamma$-ray fluxes 
   are given by  
\begin{align}
    \frac{{\rm d} 
    N_{\nu/\gamma}}{{\rm d}E_{\nu/\gamma}}
   &   =
    \frac{1}{4\pi d^2}
    \int {\rm d}V \ 
    \frac{{\rm d} \dot{n}_{\nu/\gamma}
    (E_{\nu/\gamma})}{{{\rm d} E_{\nu/\gamma}}}    
\end{align} 
(observed at Earth), 
 where  
\begin{align} 
\label{eq:hadronic}
    \frac{{\rm d} \dot{n}_{\nu/\gamma}
    (E_{\nu/\gamma})}{{{\rm d} E_{\nu/\gamma}}} 
     & =
    c\;\! n_{\rm H}
    \int_{m_{\rm p}}^\infty  
    \frac{{\rm d} E_{\rm p}}{E_{\rm p}} \ 
    F_{\nu/\gamma}
    \left(
    \frac{E_{\nu/\gamma}}{E_{\rm p}},E{_{\rm p}}
    \right)
    \sigma_{\rm pp}(E_{\rm p}) \ 
    n_{\rm CRp}(E_{\rm p}, r) 
    \ .  
 \end{align}
Here, $F_{\nu/\gamma}$ is the 
    spectrum of the secondary neutrinos 
    and $\gamma$-rays produced by a single p-p collision.
We calculated this using the
    QGSJET-II-04m hadronic model
    provided by the \texttt{AAfragpy} package
    \cite{Kachelriess19CPC, Koldobskiy21PRD}.


\subsection{Central Molecular Zone}  

The observed neutrino and $\gamma$-ray fluxes 
  are linearly dependent 
  on the number density 
  of target baryons 
  (see Equation~\ref{eq:hadronic}), 
  which are predominantly Hydrogen located in the CMZ. 
Ideally, we would use a 3D number density profile to capture the distribution of CMZ baryons. However, this is 
  not directly observable.
Instead, we adopt  
  a line-of-sight Hydrogen column density, 
  which can be derived from infrared (IR) or radio observations,
  to compute a map of 
  neutrino and $\gamma$-ray intensities.   
The rationale behind this is that 
 both neutrinos and $\gamma$-rays  
 do not suffer significant attenuation when  
 propagating from the GC region  
 to reach the Earth. 
CS (carbon monosulfide) and dust emission 
  are often used as mass tracers for molecular clouds (MCs),     
  but they give  
   substantially different baryon distribution profiles  
   in the CMZ  
    \citep[cf.][]{Tsuboi99ApJS, Molinari11ApJL}. 
We therefore consider 
  both CS and dust observations,  
  taken from
    \cite{Tsuboi99ApJS} and \cite{Molinari11ApJL},
    respectively, 
    to derive Hydrogen column density maps.  
The region within
    359.4$^\circ$ to 0.95$^\circ$ in Galactic longitude and 
    $-0.25^\circ$ to $+0.15^\circ$ in Galactic latitude sufficiently
    covers the densest structure of CMZ.  
    Integration over this region 
   yields a total gas mass of $\sim7\times10^7\,$M$_\odot$. 


\section{Results and discussion}

\begin{figure}[t]
    \centering
    \includegraphics[width=0.675\textwidth]{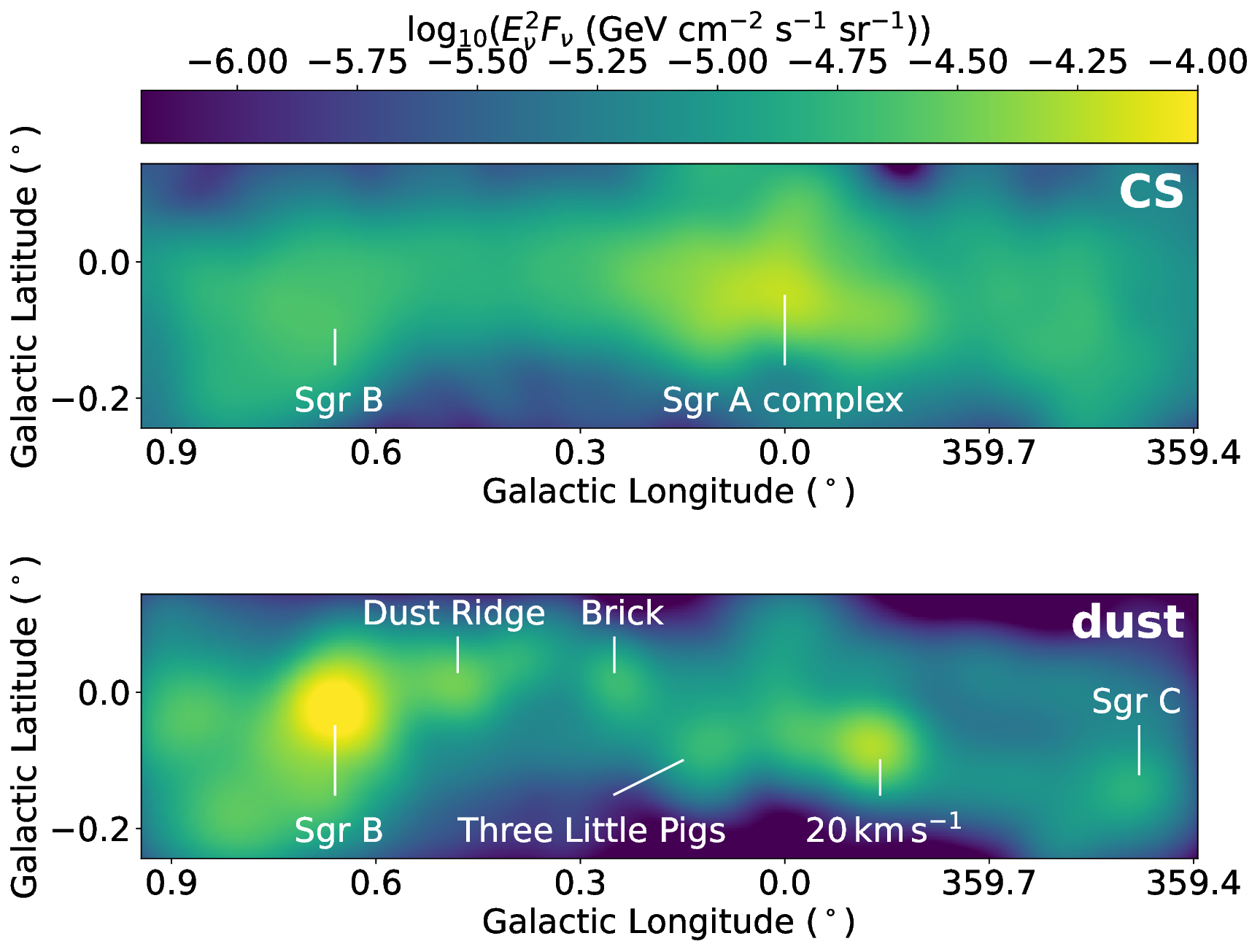}
    \caption{All-flavor neutrino surface brightness
    images at $E_\nu = 100\,{\rm TeV}$  for
    $\Gamma = -1$, 
    $N_{\rm MSP} = 5\times10^3$
    and $f_{\rm p} = 1\%$.
The effective angular resolution is $0.077^\circ$,
    which is the same as the angular resolution of the 
    CS observations \citep{Tsuboi99ApJS}.
    The panels show the results
 using CS and dust emission as the mass tracer, respectively.
    Some known MCs are marked, with others listed in Refs. \cite{Tsuboi11PASJ, Battersby20ApJS}.}
\label{fig:image}
\end{figure}

We adopt a distance of $d=8.28\,$kpc 
    from Earth to the GC
    \cite{GRAVITY21A&A} 
     and a diffusion coefficient 
     $D(E_{\rm p})  = 2.24 \times 10^{28} \ 
    {\rm cm}^2\,{\rm s}^{-1}
    \left(
    {E_{\rm p}}/{1\,{\rm GeV}}
    \right)^{\alpha}$ \citep{Gaggero15PRD}  ,
    where $\alpha = 0.21$
      for the transport of CRs near the GC.   
Two groups of MSPs, with 
    $(B,R_{\rm ns}, P_{\rm s})
    = (10^8\,{\rm G},
    10\,{\rm km},
    1\,{\rm ms})$
    and 
    $(B,R_{\rm ns}, P_{\rm s})
    = (10^{10}\,{\rm G},
    10\,{\rm km},
    10\,{\rm ms})$,  
    are considered in our calculations.  
These have the same spin-down luminosity but a 
    different maximum accelerated CR proton energy.
They are uniformly distributed within
    a galacto-centric radius of 20\,pc.
The energy spectral indices of the CR protons that they produce are    
    $\Gamma = -1$, $-1.5$ and $-2$.

Figure~\ref{fig:image} shows the all-flavour 
    neutrino surface brightness
    images at $E_\nu = 100\,{\rm TeV}$ 
    for the case with  
    $N_{\rm MSP} = 5\times10^3$,
    $f_{\rm p} = 1\%$,
    and MSP CR energy spectral index $\Gamma=-1$.  
In generating the neutrino surface brightness images, 
  we also considered other choices of CR spectral properties.  
We found that 
  the choices of $\Gamma$ and $E_\nu$ 
  do not alter the presence or absence of the locally peaked emission in the neutrino maps, 
  which correspond to the region with a high concentration of baryons.  
However, they do change the brightness normalisation, 
  implying there is a degeneracy of the MSP CR spectra parameters, MSP number density, and the efficiency of converting MSP 
  rotational energy to CR luminosity. 
The effective angular resolution of the images is $0.077^\circ$. 
Figure~\ref{fig:image} indicates that the features shown in the two panels 
  could be resolved at a spatial resolution of about 
  $0.2^\circ$, or even $0.3^\circ$, 
  which would be achievable with 
  new observatories located in the northern
  hemisphere, 
  e.g. KM3NeT, Baikal-GVD and P-ONE.

By comparing the two panels 
  in Figure~\ref{fig:image} we may draw two qualitative conclusions.  
First, modelling of the baryon column density 
  derived from CS emission or dust emission 
  gives different features in the neutrino surface brightness map. 
A direct consequence of this is that 
  CRs in the GC region are baryon ``torches", 
  and neutrino imaging can be used as a means  
  to assess the reliability of 
  different mass tracers of MCs in the GC. 
Second, integrating the surface brightness over the image 
  gives the emission power of neutrinos and, hence, yields 
  a constraint on the power of the CRs that have produced them. 
This gives us some information about  
 the spectral properties of MSP CRs, 
  specified by $\Gamma$ and $E_{\rm p,max}$,  
  the number density of MSPs,
  and the efficiency with which MSPs can produce CRs ($f_{\rm p}$). 
A strong constraint cannot be drawn due to  degeneracies between these parameters 
  and the convolution between some of them.

\begin{figure}
    \centering
    \includegraphics[width=0.48\textwidth]{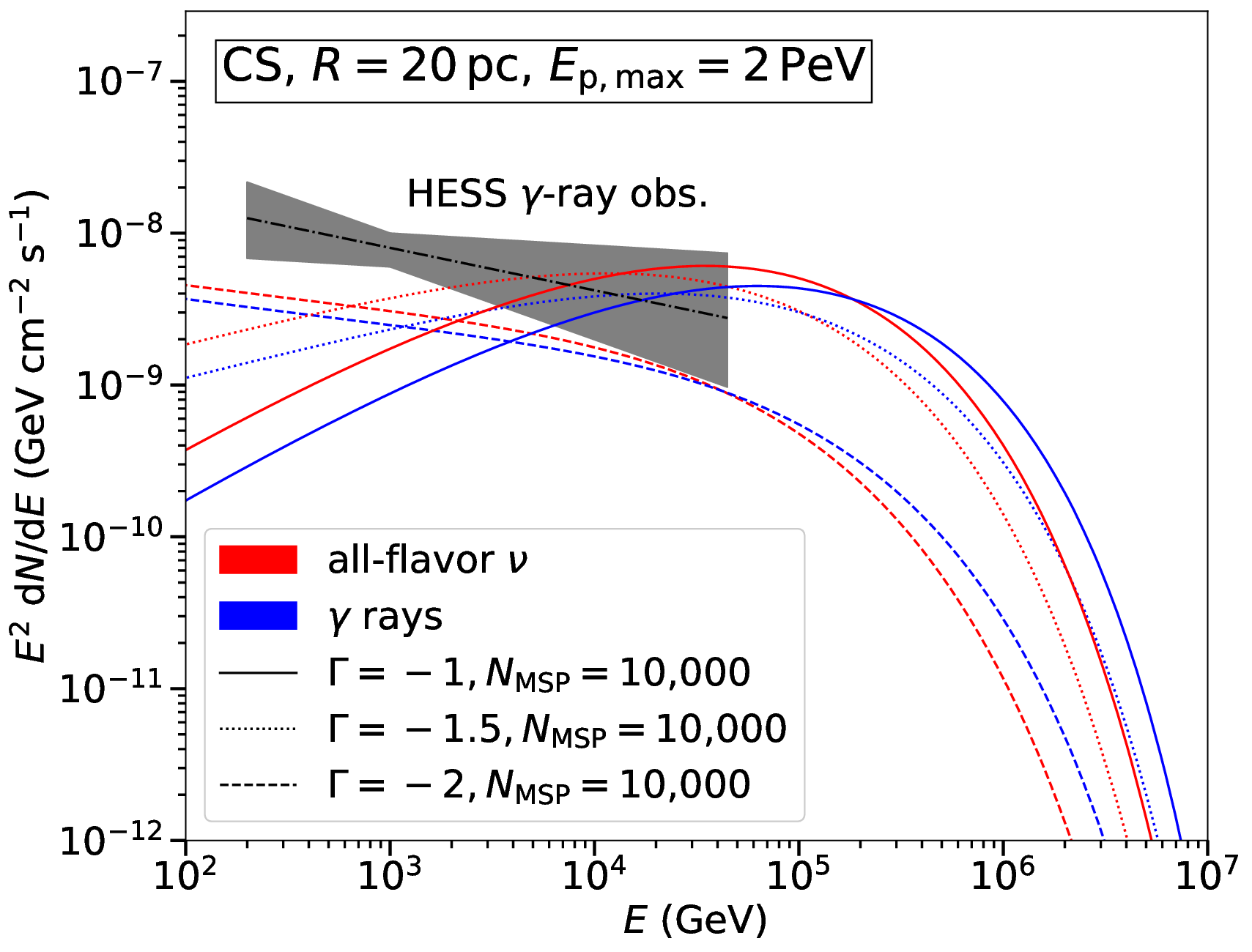}
    \includegraphics[width=0.48\textwidth]{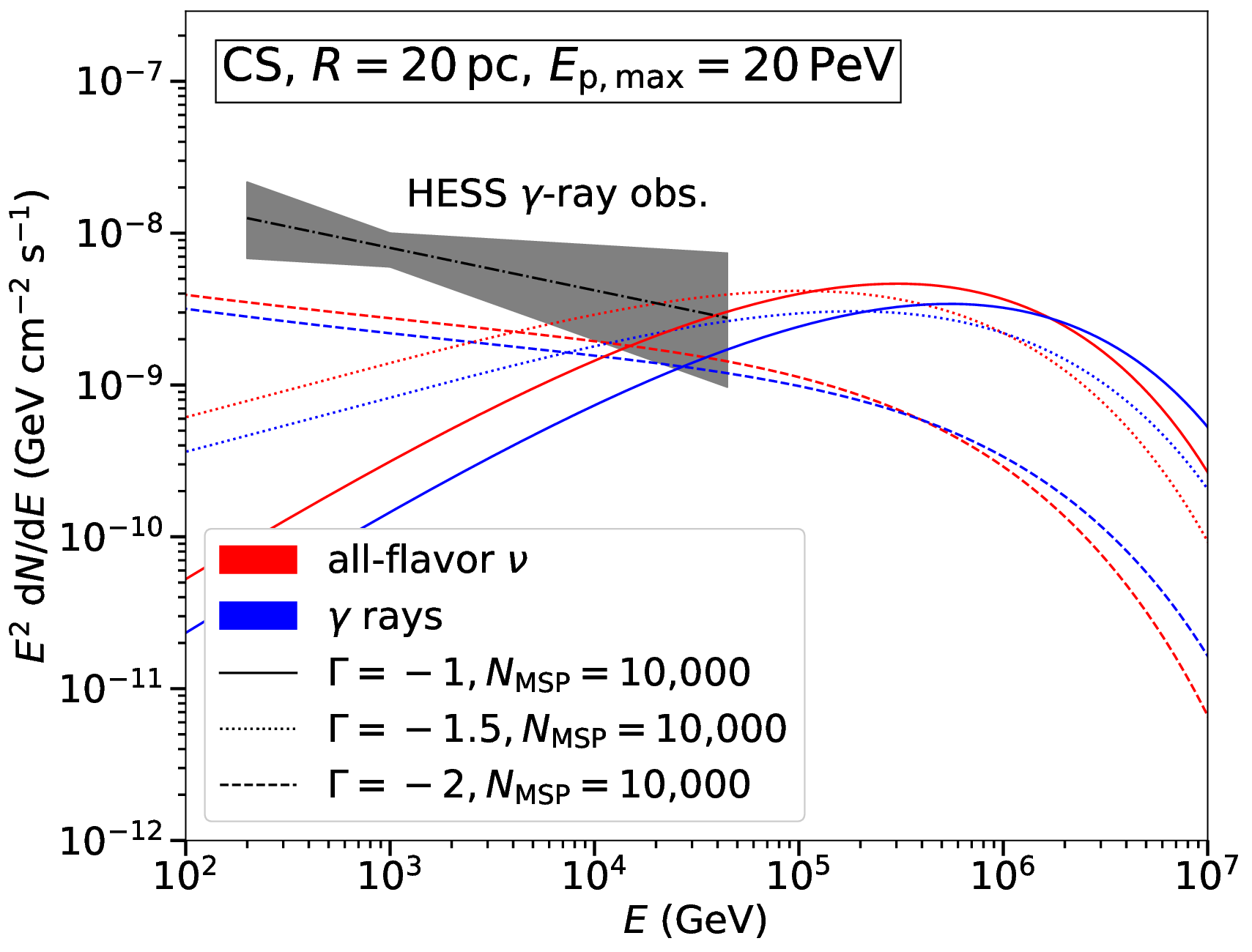} \\ 
    \includegraphics[width=0.48\textwidth]{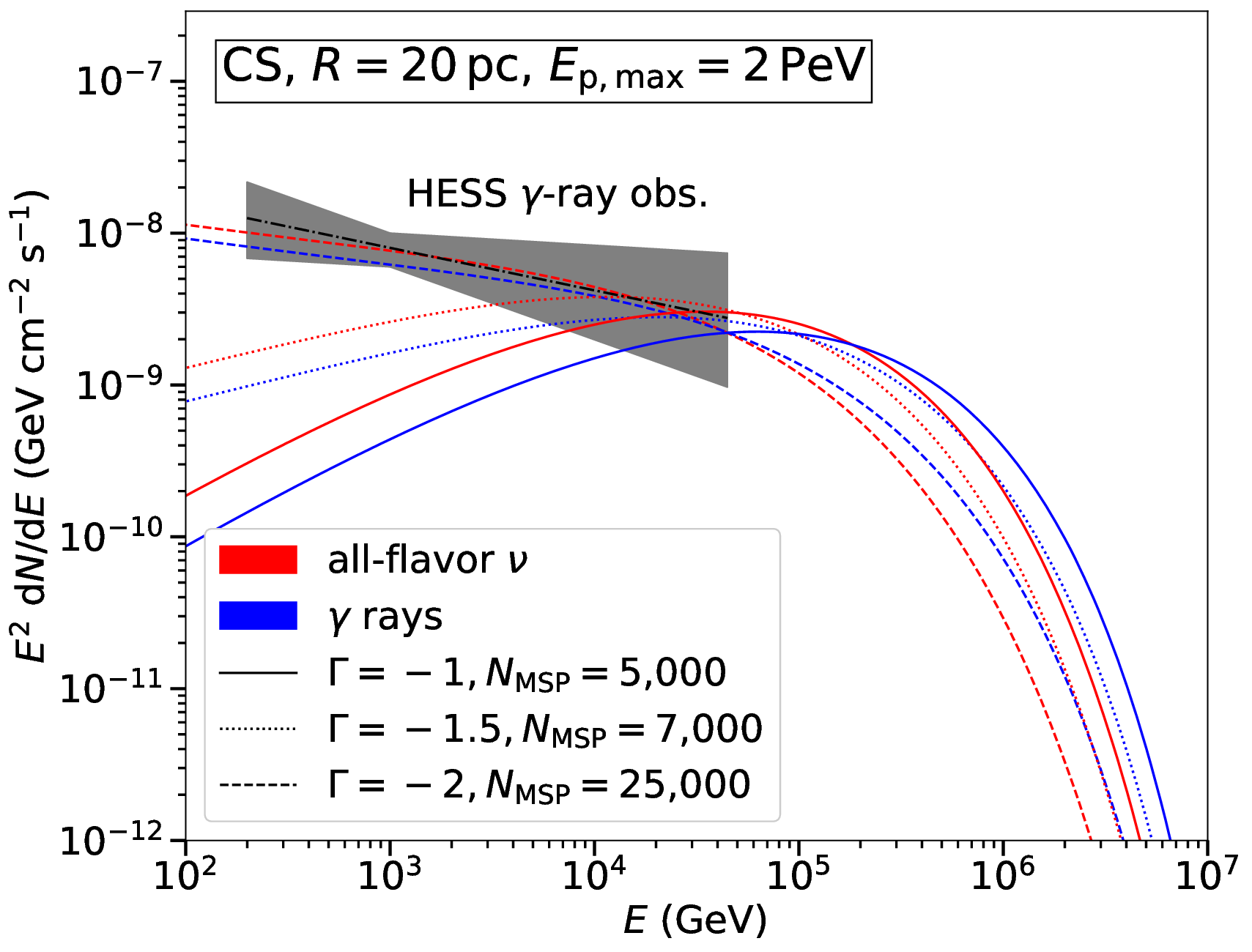}
    \includegraphics[width=0.48\textwidth]{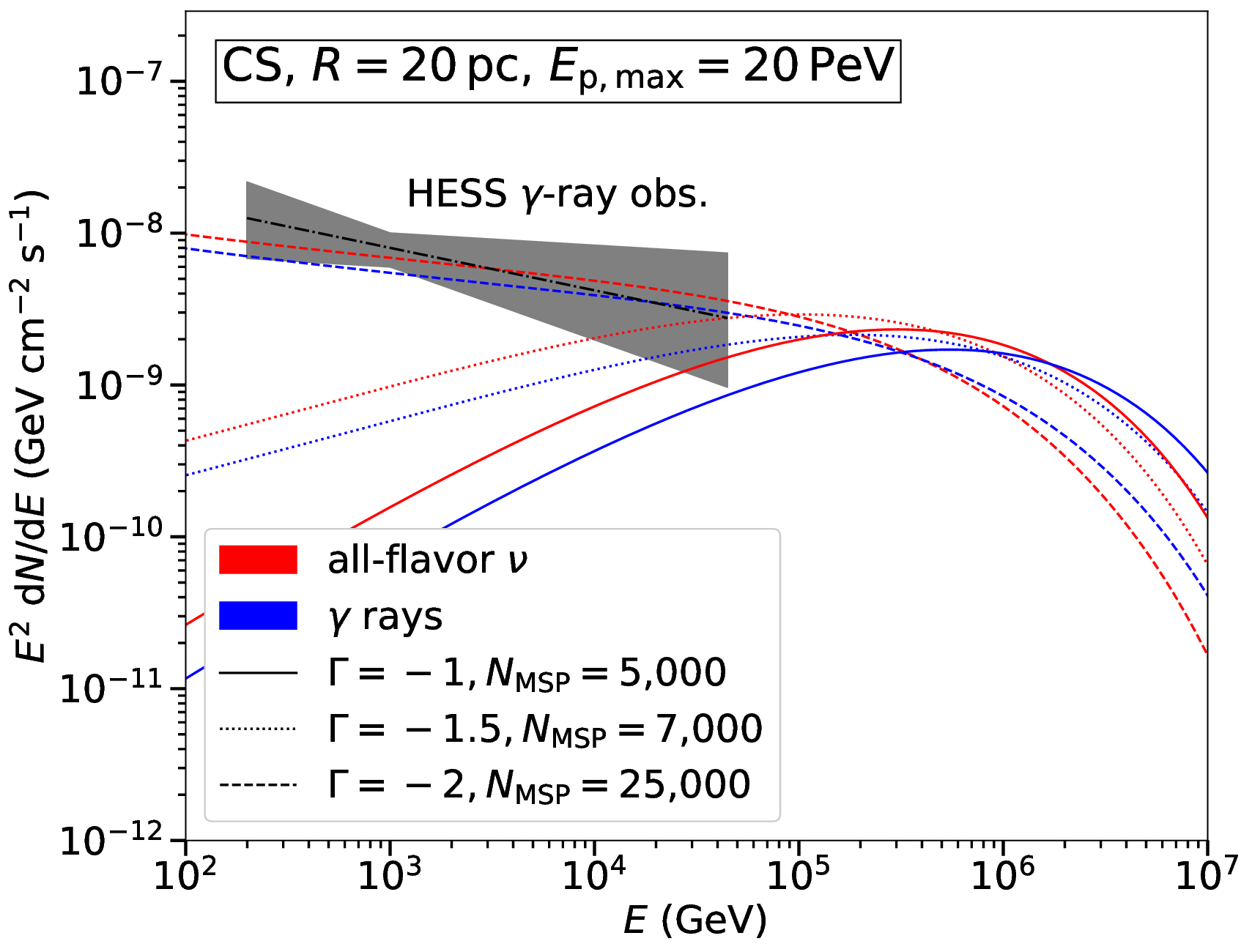}
    \caption{The computed total spectra of 
         all-flavor neutrino and $\gamma$-ray 
        emission from  the entire CMZ within a galacto-centric radius of $R=20~{\rm pc}$, 
        using CS emission as the baryon tracer.  
     The spectrum of the $\gamma$-rays from the GC 
        detected by H.E.S.S. \citep{HESS18A&A}  
       is also shown as a dotted dash line, 
       with the uncertainties represented by the grey band.   
    In the upper panels,  
    $N_{\rm MSP}$ is set to be the same for all the models. 
    In the lower panels, 
    the normalisation for $N_{\rm MSP}$ is chosen 
    such that 
    the computed specific $\gamma$-ray flux  
    matches that derived from H.E.S.S.~observations. 
    The acceleration efficiency $f_{\rm p}$ is set to be 
    1 per cent for all our calculations.
    }
\label{fig:total}
\end{figure}

Figure~\ref{fig:total} shows the spectra 
  of the total all-flavor neutrino and $\gamma$-ray emission  
  from the entire CMZ.   
The upper panels show how their spectra change with
    $\Gamma$ by fixing the normalisation
    $f_{\rm p}N_{\rm MSP}$.
We then adjust the population $N_{\rm MSP}$ for different $\Gamma$ to match
 H.E.S.S. observations \citep{HESS18A&A}, 
    which is shown in the lower panels 
    of Figure~\ref{fig:total}.
The $\gamma$-ray flux derived from our models should not exceed the H.E.S.S. observation, which constrains the
  possible population of MSPs.
For $\Gamma = -1$,
    we obtained a conservative upper limit
    of $N_{\rm MSP}<10{,}000$, assuming
    that $f_{\rm p}=1\%$.
In all the models considered, 
   the all-flavour neutrino specific flux tracks the $\gamma$-ray specific flux closely. 
Similar brightness features are therefore expected 
  to be present in the neutrino and $\gamma$-ray 
  surface brightness maps
  if there is no leptonic $\gamma$-ray emission.
This implies that spectral 
   imaging neutrino observations  
   allow the properties of CR accelerators to be probed, and their population to be constrained (if adopting a specific candidate accelerator, such as GC MSPs). 

\section{Conclusion}

We consider MSPs as particle accelerators 
  whose CRs interact with the baryons 
  in the CMZ of the GC through hadronic processes. 
We calculate the neutrino and $\gamma$-ray  emission 
  resulting from these hadronic processes, and show neutrino images of the GC  
  using CS and dust emission 
  as tracers of target baryons  
  in the CMZ. We demonstrate that 
  neutrino imaging could be used 
  to determine the baryon distribution in the GC. 
We compute the specific fluxes of neutrino and $\gamma$-ray 
  emission for the case of CS emission as the baryon tracer, 
  subject to the limits set by H.E.S.S. observations.  
We obtain a conservative upper limit 
   of $N_{\rm MSP} < 10{,}000$ for the GC MSP population,  
   with a CR energy spectral index $\Gamma = -1$
   and efficiency of $f_{\rm p} = 1$ per cent, 
   by converting the rotational power of pulsars to CR power.   
This limit varies with the model parameters, 
   and is inversely proportional to $f_{\rm p}$. 
The sizes of MSP populations that we obtain 
   are consistent with those required to account for 
    the GeV $\gamma$-ray excess in GC. 
The results obtained in this work are a benchmark for future neutrino detectors that will observe
    the GC at high angular resolution ($<1^\circ$).

\acknowledgments 
PCWL is supported by the UCL Graduate Research Scholarship and Overseas Research Scholarship. PCWL, MA, BC and KW are supported by the UCL Cosmoparticle Initiative.
ERO is an international research fellow under the Postdoctoral Fellowship of the Japan Society for the Promotion
of Science (JSPS), supported by JSPS KAKENHI Grant Number JP22F22327. 

\bibliographystyle{JHEP}
\bibliography{neutrino}

\providecommand{\noopsort}[1]{}\providecommand{\singleletter}[1]{#1}%

\providecommand{\href}[2]{#2}\begingroup\raggedright\begin{thebibliography}{10}

\bibitem{Pfahl2004ApJ}
E.~{Pfahl} and A.~{Loeb}, \emph{{Probing the Spacetime around Sagittarius A*
  with Radio Pulsars}}, \href{https://doi.org/10.1086/423975}{\emph{\apj}
  {\bfseries 615} (2004) 253}
  [\href{https://arxiv.org/abs/astro-ph/0309744}{{\ttfamily
  astro-ph/0309744}}].

\bibitem{Wharton12ApJ}
R.S.~{Wharton}, S.~{Chatterjee}, J.M.~{Cordes}, J.S.~{Deneva} and
  T.J.W.~{Lazio}, \emph{{Multiwavelength Constraints on Pulsar Populations in
  the Galactic Center}},
  \href{https://doi.org/10.1088/0004-637X/753/2/108}{\emph{\apj} {\bfseries
  753} (2012) 108} [\href{https://arxiv.org/abs/1111.4216}{{\ttfamily
  1111.4216}}].

\bibitem{Buhler+14}
R.~{B{\"u}hler} and R.~{Blandford}, \emph{{The surprising Crab pulsar and its
  nebula: a review}},
  \href{https://doi.org/10.1088/0034-4885/77/6/066901}{\emph{Reports on
  Progress in Physics} {\bfseries 77} (2014) 066901}
  [\href{https://arxiv.org/abs/1309.7046}{{\ttfamily 1309.7046}}].

\bibitem{Blasi00ApJL}
P.~{Blasi}, R.I.~{Epstein} and A.V.~{Olinto}, \emph{{Ultra-High-Energy Cosmic
  Rays from Young Neutron Star Winds}},
  \href{https://doi.org/10.1086/312626}{\emph{\apjl} {\bfseries 533} (2000)
  L123} [\href{https://arxiv.org/abs/astro-ph/9912240}{{\ttfamily
  astro-ph/9912240}}].

\bibitem{Arons03ApJ}
J.~{Arons}, \emph{{Magnetars in the Metagalaxy: An Origin for Ultra-High-Energy
  Cosmic Rays in the Nearby Universe}},
  \href{https://doi.org/10.1086/374776}{\emph{\apj} {\bfseries 589} (2003) 871}
  [\href{https://arxiv.org/abs/astro-ph/0208444}{{\ttfamily
  astro-ph/0208444}}].

\bibitem{Guepin20A&A}
C.~{Gu{\'e}pin}, B.~{Cerutti} and K.~{Kotera}, \emph{{Proton acceleration in
  pulsar magnetospheres}},
  \href{https://doi.org/10.1051/0004-6361/201936816}{\emph{\aap} {\bfseries
  635} (2020) A138} [\href{https://arxiv.org/abs/1910.11387}{{\ttfamily
  1910.11387}}].

\bibitem{HESS16Nat}
{H.E.S.S. Collaboration}, A.~{Abramowski}, F.~{Aharonian}, F.A.~{Benkhali},
  A.G.~{Akhperjanian}, E.O.~{Ang{\"u}ner} et~al., \emph{{Acceleration of
  petaelectronvolt protons in the Galactic Centre}},
  \href{https://doi.org/10.1038/nature17147}{\emph{\nat} {\bfseries 531} (2016)
  476} [\href{https://arxiv.org/abs/1603.07730}{{\ttfamily 1603.07730}}].

\bibitem{HESS18A&A}
{H.~E.~S.~S. Collaboration}, H.~{Abdalla}, A.~{Abramowski}, F.~{Aharonian},
  F.~{Ait Benkhali}, A.G.~{Akhperjanian} et~al., \emph{{Characterising the VHE
  diffuse emission in the central 200 parsecs of our Galaxy with H.E.S.S.}},
  \href{https://doi.org/10.1051/0004-6361/201730824}{\emph{\aap} {\bfseries
  612} (2018) A9} [\href{https://arxiv.org/abs/1706.04535}{{\ttfamily
  1706.04535}}].

\bibitem{MAGIC20A&A}
{MAGIC Collaboration}, V.A.~{Acciari}, S.~{Ansoldi}, L.A.~{Antonelli},
  A.~{Arbet Engels}, D.~{Baack} et~al., \emph{{MAGIC observations of the
  diffuse {\ensuremath{\gamma}}-ray emission in the vicinity of the Galactic
  center}}, \href{https://doi.org/10.1051/0004-6361/201936896}{\emph{\aap}
  {\bfseries 642} (2020) A190}
  [\href{https://arxiv.org/abs/2006.00623}{{\ttfamily 2006.00623}}].

\bibitem{Rogers+22}
F.~{Rogers}, S.~{Zhang}, K.~{Perez}, M.~{Clavel} and A.~{Taylor}, \emph{{New
  Constraints on Cosmic Particle Populations at the Galactic Center Using X-Ray
  Observations of the Molecular Cloud Sagittarius B2}},
  \href{https://doi.org/10.3847/1538-4357/ac7717}{\emph{\apj} {\bfseries 934}
  (2022) 19}.

\bibitem{Guo12ApJ}
F.~{Guo} and W.G.~{Mathews}, \emph{{The Fermi Bubbles. I. Possible Evidence for
  Recent AGN Jet Activity in the Galaxy}},
  \href{https://doi.org/10.1088/0004-637X/756/2/181}{\emph{\apj} {\bfseries
  756} (2012) 181} [\href{https://arxiv.org/abs/1103.0055}{{\ttfamily
  1103.0055}}].

\bibitem{Yang22NatAst}
H.Y.K.~{Yang}, M.~{Ruszkowski} and E.G.~{Zweibel}, \emph{{Fermi and eROSITA
  bubbles as relics of the past activity of the Galaxy's central black hole}},
  \href{https://doi.org/10.1038/s41550-022-01618-x}{\emph{Nature Astronomy}
  {\bfseries 6} (2022) 584}.

\bibitem{Marcowith18MNRAS}
A.~{Marcowith}, V.V.~{Dwarkadas}, M.~{Renaud}, V.~{Tatischeff} and
  G.~{Giacinti}, \emph{{Core-collapse supernovae as cosmic ray sources}},
  \href{https://doi.org/10.1093/mnras/sty1743}{\emph{\mnras} {\bfseries 479}
  (2018) 4470} [\href{https://arxiv.org/abs/1806.09700}{{\ttfamily
  1806.09700}}].

\bibitem{Morlino21MNRAS}
G.~{Morlino}, P.~{Blasi}, E.~{Peretti} and P.~{Cristofari}, \emph{{Particle
  acceleration in winds of star clusters}},
  \href{https://doi.org/10.1093/mnras/stab690}{\emph{\mnras} {\bfseries 504}
  (2021) 6096} [\href{https://arxiv.org/abs/2102.09217}{{\ttfamily
  2102.09217}}].

\bibitem{IceCube2023Sci}
IceCube{~}Collaboration, \emph{{Observation of high-energy neutrinos from the
  Galactic plane}},
  \href{https://doi.org/doi.org/10.1126/science.adc9818}{\emph{Science}
  {\bfseries 380} (2023) eaat1338}.

\bibitem{Bartels16PRL}
R.~{Bartels}, S.~{Krishnamurthy} and C.~{Weniger}, \emph{{Strong Support for
  the Millisecond Pulsar Origin of the Galactic Center GeV Excess}},
  \href{https://doi.org/10.1103/PhysRevLett.116.051102}{\emph{\prl} {\bfseries
  116} (2016) 051102}.

\bibitem{Kachelriess19CPC}
M.~{Kachelrie{\ss}}, I.V.~{Moskalenko} and S.~{Ostapchenko}, \emph{{AAfrag:
  Interpolation routines for Monte Carlo results on secondary production in
  proton-proton, proton-nucleus and nucleus-nucleus interactions}},
  \href{https://doi.org/10.1016/j.cpc.2019.08.001}{\emph{Computer Physics
  Communications} {\bfseries 245} (2019) 106846}.

\bibitem{Koldobskiy21PRD}
S.~{Koldobskiy}, M.~{Kachelrie{\ss}}, A.~{Lskavyan}, A.~{Neronov},
  S.~{Ostapchenko} and D.V.~{Semikoz}, \emph{{Energy spectra of secondaries in
  proton-proton interactions}},
  \href{https://doi.org/10.1103/PhysRevD.104.123027}{\emph{\prd} {\bfseries
  104} (2021) 123027} [\href{https://arxiv.org/abs/2110.00496}{{\ttfamily
  2110.00496}}].

\bibitem{Tsuboi99ApJS}
M.~{Tsuboi}, T.~{Handa} and N.~{Ukita}, \emph{{Dense Molecular Clouds in the
  Galactic Center Region. I. Observations and Data}},
  \href{https://doi.org/10.1086/313165}{\emph{\apjs} {\bfseries 120} (1999) 1}.

\bibitem{Molinari11ApJL}
S.~{Molinari}, J.~{Bally}, A.~{Noriega-Crespo}, M.~{Compi{\`e}gne},
  J.P.~{Bernard}, D.~{Paradis} et~al., \emph{{A 100 pc Elliptical and Twisted
  Ring of Cold and Dense Molecular Clouds Revealed by Herschel Around the
  Galactic Center}},
  \href{https://doi.org/10.1088/2041-8205/735/2/L33}{\emph{\apjl} {\bfseries
  735} (2011) L33} [\href{https://arxiv.org/abs/1105.5486}{{\ttfamily
  1105.5486}}].

\bibitem{Tsuboi11PASJ}
M.~{Tsuboi}, K.-I.~{Tadaki}, A.~{Miyazaki} and T.~{Handa}, \emph{{Sagittarius A
  Molecular Cloud Complex in H$^{13}$CO$^{+}$ and Thermal SiO Emission Lines}},
  \href{https://doi.org/10.1093/pasj/63.4.763}{\emph{\pasj} {\bfseries 63}
  (2011) 763}.

\bibitem{Battersby20ApJS}
C.~{Battersby}, E.~{Keto}, D.~{Walker}, A.~{Barnes}, D.~{Callanan},
  A.~{Ginsburg} et~al., \emph{{CMZoom: Survey Overview and First Data
  Release}}, \href{https://doi.org/10.3847/1538-4365/aba18e}{\emph{\apjs}
  {\bfseries 249} (2020) 35}
  [\href{https://arxiv.org/abs/2007.05023}{{\ttfamily 2007.05023}}].

\bibitem{GRAVITY21A&A}
{GRAVITY Collaboration}, R.~{Abuter}, A.~{Amorim}, M.~{Baub{\"o}ck},
  J.P.~{Berger}, H.~{Bonnet} et~al., \emph{{Improved GRAVITY astrometric
  accuracy from modeling optical aberrations}},
  \href{https://doi.org/10.1051/0004-6361/202040208}{\emph{\aap} {\bfseries
  647} (2021) A59} [\href{https://arxiv.org/abs/2101.12098}{{\ttfamily
  2101.12098}}].

\bibitem{Gaggero15PRD}
D.~{Gaggero}, A.~{Urbano}, M.~{Valli} and P.~{Ullio}, \emph{{Gamma-ray sky
  points to radial gradients in cosmic-ray transport}},
  \href{https://doi.org/10.1103/PhysRevD.91.083012}{\emph{\prd} {\bfseries 91}
  (2015) 083012} [\href{https://arxiv.org/abs/1411.7623}{{\ttfamily
  1411.7623}}].

\end{thebibliography}\endgroup





%
%
%

\end{document}